\documentclass[10pt]{article}

\usepackage{amsmath,amssymb,dsfont,slashed,color,hyperref}
\usepackage{authblk}
\usepackage[toc,page]{appendix}
\usepackage{graphicx,caption,subcaption,tikz}

\newlength\imagewidth
\newlength\imagescale

\newcommand{\half}{\frac{1}{2}}

\newcommand{\QQ}{\mathds{Q}}
\newcommand{\QQb}{\overline{\mathds{Q}}}

\newcommand{\se}{\slashed{e}}

\newcommand{\lD}{\overleftarrow{D}}

\newcommand{\oD}{\mathring{D}}

\newcommand{\psibar}{{\overline{\psi}}}
\newcommand{\chibar}{{\overline{\chi}}}

\newcommand{\uantof}{Departamento de F\'isica, Universidad de Antofagasta, Aptdo. 02800, Chile}

\begin{document}

\title{Spinor solutions of a Chern-Simons model for the superconformal algebra}
\author[1]{Pedro D. Alvarez \thanks{E-mail: \href{mailto:pedro.alvarez@uantof.cl}{\nolinkurl{pedro.alvarez@uantof.cl}}}}

\author[1]{Juan Ortiz \thanks{E-mail: \href{mailto:juan.ortiz.camacho@ua.cl}{\nolinkurl{juan.ortiz.camacho@ua.cl}}}}
\affil[1]{\uantof}

\date{}
\maketitle


\begin{abstract}

We present analytical solutions for homogenous and isotropic spaces of the supersymmetric Chern-Simons model with matter in the adjoint representation. The configurations that we found correspond to a gravitating spinor content and torsion is also present. The spinor behaves like dark energy in the sense that drives an exponential expansion. The solution found can be seen as an anisotropic fluid.

\end{abstract}

\maketitle

\section{Introduction}\label{int}

Three-dimensional theories of gravity have a very rich structure of black hole solutions with matter: polytropic stars from \cite{Sa:1999yf}, rigidly rotating perfect fluid stars  \cite{Gundlach:2020ovt}, low mass strange stars based on the Heintzmann ansatz \cite{Murshid:2021iaq}, circular thin shells in 2+1 $F(R)$ gravity \cite{Eiroa:2020dip}, exact solutions in (2 + 1)-dimensional anti-de Sitter space-time admitting a linear or non-linear equation of state \cite{Banerjee:2014mwa} and references therein.

Gravitational collapse is a highly dissipating process. In the dynamics of dissipation during the collapse process, an important role is played by the pressure anisotropy, which is the difference of pressures generated in a system between radial and tangential directions. In relativistic astrophysics, the consideration of anisotropic stress in a collapsing system originates from various factors. In \cite{Das:2020rfz}, the authors studied the gravitational collapse of a spherically symmetric anisotropic relativistic star within Einstein’s theory of gravity. Assuming an interacting equation of state with anisotropic configuration of strange matter, in \cite{Tangphati:2021tcy}, the authors focused on the properties of quark stars, which are excellent laboratories for matter under extreme conditions.

Anisotropy in cosmological solutions has also been studied by McManus and Coley \cite{McManus:1994ys}, Giovannini \cite{Giovannini:1998rh}. Relative motion as a source of anisotropy in multi-fluid systems was suggested long ago by Jeans \cite{Jeans1922:10.1093/mnras/82.3.122} and in the 80s Letelier showed that two perfect fluids in relative motion can be described as a system with anisotropic pressures with a standard two-fluid stress energy form \cite{Letelier:1980mxb}.

The interacting system of gravity and spinors can be studied, but exact solutions are hard to find. For this reason it is convenient to search for solutions in simple systems. In \cite{Cianci:2015pba}, the authors found solutions for a neutral spinor field. In \cite{Cianci:2016pvd}, the authors considered the Einstein-Dirac field equations describing a self-gravitating massive neutrino, and looked for axially-symmetric exact solutions. It was found some solutions with critical features, such as the fact that the space-time curvature turns out to be flat and the spinor field gives rise to a vanishing bi-linear scalar with non-vanishing bi-linear pseudo-scalar. In addition to this the solutions of \cite{Cianci:2016pvd}, have non-zero momentum density vector such that is a covariantly constant null vector. The presence of the non-diagonal components of energy-momentum tensor of the spinor field leads to some severe restrictions on the spinor field itself \cite{Mei:2011mh}. Since spinor fields are sources of the gravitational fields the system as a whole possesses solutions only in case of some additional conditions on metric functions \cite{Saha:2018ufp}. See also recently solutions in G\"odel-type geometries \cite{Ahmed:2019vmv}. In \cite{Saha:2019ztr}, there were found numerical solutions for a non-minimally coupled, gravitating spinor in a FRW background.

Let us remark in passing that there has been a lot of interest in the study of Weyl fermions because of the discovery of such a particles inside tantalum arsenide (a synthetic metallic crystal) \cite{Xu:2015cga}. This finding can be used to avoid electron backscattering, therefore improving the efficiency of electronic devices. Weyl fermions are also important from the fundamental point of view since it is still open the possibility that neutrinos may oscillate because of intrinsic dynamics \cite{Fabbri:2015jaa}.

In the present paper we consider spinors in the adjoint representation of supersymetry,
\begin{equation}
 \QQb\se\psi+\psibar\se\QQ \subset \mathbb{A}\label{connectionintro}
\end{equation}
where $\se =e^a\gamma_a$ and $e^a = e^a{}_\mu dx^\mu$ is a set of orthornormal frames. The composite field $e^a \gamma^a \psi$ are fermonic one-forms and they are killed by the spin 3/2 projector. Such embedding of spin 1/2 fields in the adjoint representation of the superalgebra is the so called `unconventional supersymmetry' \cite{Alvarez:2011gd,Alvarez:2013tga,Alvarez:2015bva,Alvarez:2020qmy,Alvarez:2020izs,Alvarez:2021zhh,Alvarez:2021zsw}. In 2+1 dimensions the action is based on the Chern-Simons term and in \cite{Guevara:2016rbl} the authors showed that the only propagating local degrees of freedom are the fermions. In \cite{Alvarez:2014uda} there were found geometries for the three dimensional black hole with torsion, see also \cite{Alvarez:2015bva} for nontrivial $SU(2)$ solutions.

The paper is organized as follows. In sect.~\ref{sec:1}, we define the model. In sect. \ref{fieldeqs}, we present the field equations and we study the constraints on the spinor bilinears coming from the on-shell restriction of the Bianchi-identities. In sec. \ref{spinoreqs}, we solve the spinor equations and we discuss the physical meaning of the solution. In sect.~\ref{sec:conclu}, we summarize our results and provide some ideas for the future developments.

\section{SUSY in the adjoint representation}\label{sec:1}

Let us consider a model with matter in the adjoint representation for a $su(2|2)\times su(2|2)$ gauge potential
\begin{equation}
 \mathbb{A}=\mathbb{A}_++\mathbb{A}_-\,,
\end{equation}
where
\begin{align}
 \mathbb{A}_+&=\half \omega^{ab}_+ \mathbb{J}^+_a+A^I_+ \mathbb{T}^+_I+\overline{\mathbb{Q}}^{+i}\se\psi_i+\overline{\psi}^i\se\mathbb{Q}^+_i+b_+ \mathbb{Z}^+\,,\label{Aplus}\\
 \mathbb{A}_-&=\half \omega^{ab}_- \mathbb{J}^-_a+A^I_- \mathbb{T}^-_I+\overline{\mathbb{Q}}^{-i}\se\chi_i+\overline{\chi}^i\se\mathbb{Q}^-_i+b_- \mathbb{Z}^-\,.\label{Aminus}
\end{align}
where the bar on the spinors is the Dirac adjoint, $\psibar \equiv i \psi^\dagger \gamma_0$. We define the action by the Achucarro-Townsend term \cite{Achucarro:1986uwr},
\begin{equation}\label{action}
 S=2\ell\int \left(L_\text{CS}(\mathbb{A}_+)-L_\text{CS}(\mathbb{A}_-)\right)\,.
\end{equation}
where
\begin{equation}
L_\text{CS}(\mathbb{A})=\frac{\kappa}{2}\langle \mathbb{A}d\mathbb{A}+\frac{2}{3}\mathbb{A}^3\rangle\,.
\end{equation}
and the angular brackets represent supertraces in the conformal superalgebra representation provided in the appendix \ref{apprep}

The Einstein-Hilbert action in three dimensions plus a cosmological constant term, with $\Lambda = -1/\ell^2$, are recovered when
\begin{align}
 \omega_\pm^{ab} & = \omega^{ab} \mp \frac{1}{\ell}\epsilon^{ab}{}_c f^c\,,\label{wpm}\\
 J^\pm_{ab} & = \frac{1}{2}(J_{ab} \mp \ell \epsilon^{ab}{}_c P_c)\,.
\end{align}
In analogy with (\ref{wpm}), let us take
\begin{equation}
 A^I_\pm=A^I\pm B^I\,,
\end{equation}
this implies that $B^I=\frac{1}{2}(A^I_+-A^I_-)$ transform as a one-form under the diagonal $SU(2)$ gauge symmetry. Expanding the action (\ref{action}), we obtain
\begin{equation}
 S=2\ell\int \left(L_\text{CS}(\omega_+)-L_\text{CS}(\omega_-)+L_\text{CS}(A_+)-L_\text{CS}(A_-)+L_\psi^+-L_\chi^-\right)\,,
\end{equation}
where
\begin{align}
L_\text{CS}(\omega_+)-L_\text{CS}(\omega_-) = \frac{\kappa}{\ell}(\half  \epsilon_{abc}f^a R^{bc}+\frac{1}{6\ell^2}\epsilon_{abc}f^a f^b f^c+\frac{1}{2}d(f^a \omega_a))\,,\\
L_\text{CS}(A_+)-L_\text{CS}(A_-) = \kappa(B^I F_I+\frac{1}{6}\epsilon_{IJK}B^I B^J B^K+\frac{1}{2}d(B^I A_I))\,,\\
L_\psi^+ - L_\chi^- = \frac{\kappa}{2}(\psibar\se )(\overleftarrow{D}^+ - D^+)(\se\psi) - \frac{\kappa}{2}(\chibar\se )(\overleftarrow{D}^- - D^-)(\se\chi)\,.
\end{align}
The spinor lagrangian can be further expanded in terms of the Lorentz covariant derivatives using
$L_\psi^\pm=L_\psi+\Lambda^\pm_\psi+\Gamma^\pm_\psi$, where
\begin{align}
 L_\psi&=\frac{\kappa}{2}(\overline{\psi}\se )(\overleftarrow{D}-D)(\se\psi)=\frac{1}{2}\overline{\psi}(\se\se D-\overleftarrow{D}\se\se )\psi-\frac{1}{2}\overline{\psi}(\se\slashed{T}+\slashed{T}\se )\psi\,,\\
 \Lambda^\pm_\psi&=\mp\frac{\kappa}{2\ell}\overline{\psi}\se\slashed{f}\se\psi\,,\\
 \Gamma^\pm_\psi&=\pm i\frac{\kappa}{2}\overline{\psi}\se B^I\sigma_I \se\psi\,.
\end{align}
The $so(1,2)\times su(2)$ covariant derivative in the spin-$1/2$ representation
\begin{eqnarray}\label{D}
 D_i^{\ j}&=&\delta^j_i d+\frac{1}{2}\delta^j_i\omega^{ab}\Sigma_{ab}-\frac{i}{2}A^I(\sigma_I)_i^{\ j}\,,\\
 \overleftarrow{D}_i^{\ j}&=&\overleftarrow{d} \delta^j_i-\frac{1}{2}\omega^{ab}\Sigma_{ab} \delta^j_i +\frac{i}{2}A^I(\sigma_I)_i^{\ j}\,,
\end{eqnarray}
and $\Omega^m\overleftarrow{d}=(-1)^m d\Omega^m$ for an $m$-form. This covariant derivative acts along the diagonal component of two copies of $so(1,2)\times su(2)$. The full Lagrangian for the spinors is
\begin{align}
L_\psi^+-L_\chi^-= &\kappa\left[\frac{1}{2}\psibar(\epsilon_{abc}e^a e^b \gamma^c D-\lD \epsilon_{abc}e^a e^b \gamma^c)\psi- e^a T_a\psibar\psi \right.\nonumber \\
&-\frac{1}{2\ell}\epsilon_{abc}e^ae^bf^c\psibar\psi-\frac{i}{2}\epsilon_{abc}e^ae^bB^I \psibar \gamma^c\sigma_I\psi\nonumber \\
&-\frac{1}{2}\chibar(\epsilon_{abc}e^a e^b \gamma^cD-\overleftarrow{D}\epsilon_{abc}e^a e^b \gamma^c)\chi+ e^a T_a\chibar\chi \nonumber\\
&\left. -\frac{1}{2\ell}\epsilon_{abc}e^ae^bf^c\chibar\chi-\frac{i}{2}\epsilon_{abc}e^ae^bB^I\chibar\gamma^c\sigma_I\chi\right]\,. \label{spinorlag}
\end{align}

\section{Field equations}\label{fieldeqs}
Variation with respect to $f^a$ and $\omega^{ab}$ give us the following eqs. of motion
\begin{align}
 &R^{ab}+\frac{1}{\ell^2}f^af^b - \Phi e^a e^b = 0\,,\label{eomf}\\
 &\frac{1}{\ell}T_f^a - \frac{1}{2}\Delta\epsilon^a_{\ bc}e^be^c = 0\,,\label{eomw}
 \end{align}
where
\begin{align}
 \Delta =& \psibar \psi - \chibar \chi\,,\\
 \Phi =& \psibar \psi + \chibar \chi\,,
\end{align}
and $T^a_f \equiv D f^a$. Note that the equation (\ref{eomf}) for $f^a$ is algebraic and in fact implies Lorentz-curvature flat geometries. All the length-dimensions of the fields are canonical in $d=3$: $[A^I_\mu]=l^{-1}$, $[f^a_\mu]=l^0=[e^a_\mu]$, $[\psi]=l^{-1}$, $[\Delta] = [\Phi] = l^{-2}$. Inspection of (\ref{eomf}) tell us that $\Phi$ has units of energy density and, if constant in time in the context of a cosmological model, could behave as dark energy. That is a peculiarity of the model, since the spin of $\psi$ is 1/2 and, see below, we will show that it is indeed posible.

Let us notice that the ansatz $f^a \sim e^a$ corresponds and $\Delta = 0$ corresponds to pure MacDowell-Mansouri gravity in $d=4$ \cite{MacDowell:1977jt} and Achucarro-Townsend gravity in $d=3$ \cite{Achucarro:1986uwr}. We can generalize the ansatz to 
\begin{equation}
 f^a=h(x)e^a\,,\label{genansatz}
\end{equation}
and look for the configurations that comply with Bianchi-identities,
\begin{align}
 D R^{ab} =& 0\,,\label{BianchiDR}\\
 D T^a_f =& R^a{}_b f^b\,.\label{BianchiDTf}
\end{align}
This strategy allow us to restrict the space of solutions by implying relations between $h$, $\Delta$ and $\Phi$ that come from the on-shell restriction of the Bianchi identities.

On the one hand, replacing (\ref{genansatz}) in (\ref{eomf}), taking the covariant derivative and using the Bianchi identity (\ref{BianchiDR}) we get
\begin{equation}
 d\left( \frac{h^2}{\ell^2}-\Phi \right) e^a e^b + \left(\frac{h^2}{\ell^2}-\Phi\right) (T^a e^b - T^b e^a) =0\,,
\end{equation}
and using $T^a$ from
\begin{equation}
 D f^a=dh\ e^a+h\ T^a \,, \label{Dfrepl}
\end{equation}
we get
\begin{equation}
 d\left( \frac{h^2}{\ell^2}-\Phi \right) e^a e^b -2h^{-1}dh \left(\frac{h^2}{\ell^2}-\Phi\right) e^a e^b =0\,.\label{h-Phi}
\end{equation}

On the other hand, taking the covariant derivative of (\ref{eomw}), using the Bianchi identity (\ref{BianchiDTf}) and again the field equation (\ref{eomf}), we get
\begin{equation}
 d\Delta \epsilon^a{}_{bc} e^b e^c +2\Delta \epsilon^a{}_{bc} T^b e^c = 0 \,. \label{Deomw}
\end{equation}
Using (\ref{Dfrepl}) to replace $T^a$ in (\ref{Deomw}), we get
\begin{equation}
 d\Delta \epsilon^a{}_{bc}e^b e^c + 2\frac{\Delta}{h}\epsilon^a{}_{bc}\left( T_f^b -dh e^b\right)e^c = 0 \,. \label{Tfafromeomw}
\end{equation}
Now, replacing $T_f^a$ from (\ref{eomw}) in (\ref{Tfafromeomw}) we get
\begin{equation}
 \frac{2\Delta}{h}dh - d\Delta = 0\,. \label{h-Delta2} 
\end{equation}
Therefore the system is not over constrained as long as the Dirac equation is compatible with conditions (\ref{h-Phi}) and (\ref{h-Delta2}).

Now, in $d = 3$, we can write the most general solution to the covariantly constant torsion condition, which is implied by (\ref{BianchiDTf}) and the field eq. (\ref{eomw}). Such solution is given by
\begin{equation}
 T_f^a = \pm \frac{\ell}{2z^2} \epsilon^a{}_{bc} f^b f^c\,, \label{torsionf3d}
\end{equation}
where $z$ is an integration constant with physical units $[z]=l$. Replacing (\ref{torsionf3d}) in (\ref{eomw}) we get the condition
\begin{equation}
 \pm\frac{h^2}{z^2} -\Delta = 0 \,. \label{zcond}
\end{equation}
Notice that (\ref{zcond}) is compatible with (\ref{h-Delta2}).

A first attempt to solve the system would be to demand the condition
\begin{equation}
 \frac{h^2}{\ell^2}-\Phi = 0\,, \label{cond2forh}
\end{equation}
which ensures that (\ref{h-Phi}) is satisfied. For latter convenience let us parametrize such condition by
\begin{equation}
 \left( \frac{z^2}{\ell^2}-\varepsilon_\Delta \theta \right) \Delta = 0\,,
\end{equation}
where $\varepsilon_\Delta = \text{sign}(\Delta)$ and
\begin{equation}
\theta \equiv \Phi /|\Delta|\,. 
\end{equation}
The solution is given by $\theta = \varepsilon_\Delta z^2/\ell^2$. Such solution, however, is not compatible with real solutions for the scale factor of the Friedmann equations. Therefore condition (\ref{cond2forh}) has to be generalized to
\begin{equation}
 \frac{h^2}{\ell^2}-\Phi = \text{const}\,, \label{cond2forh_v2}
\end{equation}
Note that in order to arrive to such conclusion we have not demanded any condition on the frames yet. Conditions (\ref{cond2forh_v2}) and (\ref{zcond}) can be satisfied when both $\Delta$ and $\Phi$ are constants.

We will show that there are exact solutions for the spinors in a FRW background with torsion. From now on we will consider FRW metric frames
\begin{equation}
 e^0 = dt\,, \quad e^i = a(t) dx^i\,, \label{FRWframes}
\end{equation}
where $a(t)$ is the scale factor. A priori we would assume a cosmological ansatz $h=h(t)$, however the condition $\Delta = \text{const}$ fixes $h=\text{const}$. The condition $\partial_t h =0$ also comes by inspecting the equation coming from (\ref{eomf}) after replacing the Lorentz curvature decomposition
\begin{equation}\label{Rab-decomp}
 R^{ab}=\mathring{R}^{ab}+\mathring{D}\kappa^{ab}+\kappa^a_{\ c}\kappa^{cb}\,,
\end{equation}
where $\mathring{R}^{ab}$ is the Riemannian two-form, where the torsionless connection is defined by $\mathring{D}f^a=0$, and $\kappa^{ab}$ is the contortion two-form defined by
\begin{equation}
 \mathring{\omega}^{ab} = \omega^{ab} + \kappa^{ab}\,.
\end{equation}
where
\begin{equation}
 \kappa^{ab} = -\frac{\ell}{2z^2} \epsilon^{ab}{}_c f^c\,.
\end{equation}
so $\mathring{D}\kappa^{ab}=0$ and $\kappa^a_{\ c}\kappa^{cb}=\frac{\ell^2}{4z^4}f^af^b$. $\Delta$ is not positive definite and therefore so we will chose
\begin{equation}
 h=z \sqrt{|\Delta |}\,,
\end{equation}
and replacing (\ref{Rab-decomp}) in (\ref{eomf}) give us
\begin{equation}\label{eq2}
  0 = \mathring{R}^{ab}-\alpha^2 |\Delta| \ e^ae^b\,,
\end{equation}
where
\begin{equation}
 \alpha^2 = \theta - \frac{\ell^4+4z^4}{4\ell^2 z^2}\,. \label{alphacond}
\end{equation}
The Friedmann equations come from using the FRW frames in (\ref{eq2}), where $\mathring{R}^{0i}=\ddot{a}dt dx^i$ and $\mathring{R}^{ij}=\dot{a}^2dx^i dx^j$,
\begin{align}
 0=&\ddot{a} - \alpha^2 |\Delta| a\,,\\
 0=&\dot{a}^2 - \alpha^2 |\Delta| a^2\,. \label{Friedmann}
\end{align}
There are exponential expanding solutions to these equations that are given by
\begin{equation}
 a(t)=a_0 \exp [\pm \alpha\sqrt{|\Delta|}(t-t_0)]\,. \label{scalefac}
\end{equation}
From (\ref{alphacond}) we see that the scale factor is real for
\begin{equation}
 \theta > \theta_0\,, \quad \text{where} \quad \theta_0 = \frac{\ell^4+4z^4}{4\ell^2 z^2}\,. \label{theta0}
\end{equation}

Before ending the section let us comment that the choice of positive sign in eq. (\ref{torsionf3d}) will imply that the physical solutions have $\Delta >0$, but solutions with $\Delta < 0$ can also be included with the choice of negative sign in eq. (\ref{torsionf3d}). For the sake of the simplicity of the discussion hereafter we will consider the positive sign in eq. (\ref{torsionf3d}) only, and therefore $\Delta >0$.

\subsection{Solution of the spinor equations}\label{spinoreqs}

The Dirac equations of the model when $B^I = 0$ are given by
\begin{align}
 0 = \gamma^\mu \oD_\mu \psi +\half |e|^{-1} \partial_\mu(|e|\gamma^\mu) \psi -\frac{3h}{2\ell}\left(\frac{\ell^2}{2z^2} + 1\right) \psi\,,\label{diracpsi}\\
 0 = \gamma^\mu \oD_\mu \chi +\half |e|^{-1} \partial_\mu(|e|\gamma^\mu) \chi -\frac{3h}{2\ell}\left(\frac{\ell^2}{2z^2} - 1\right) \chi\,,\label{diracchi}
\end{align}
where
\begin{equation}
 \mathring{D}_\mu \psi = \partial _\mu \psi +\frac{1}{2} \mathring{\omega}^{ab}{}_\mu\Sigma_{ab}\psi-\frac{i}{2} A^I{}_\mu \sigma_I \psi \,.
  \end{equation}
We will consider the flat $su(2)$ case, $A^I=0$, and only one flavor, $\psi=\psi_1$, $\psi_2=0$, $\chi=\chi_1$ and $\chi_2=0$. The strategy to solve the system given by eqs. (\ref{diracpsi}) and (\ref{diracchi}) in this paper is to treat $h$ as a constant, and check the consistency of the relations $\Delta =\text{const}$ and $\theta = \text{const} > \theta_0$ a posteriori. In terms of the spinor components we have 
\begin{equation}
\Delta = i (v_\psi(t)\overline{u_\psi(t)}- u_\psi(t)\overline{v_\psi(t)}) -i (v_\chi(t)\overline{u_\chi(t)}- u_\chi(t)\overline{v_\chi(t)})\,,\label{deltacomp}
\end{equation}
where $(u_\psi,v_\psi)=\psi$, $(u_\psi,v_\psi)=\chi$. Consistent solutions can only appear for specific values of $\Delta$ and $\theta$. The general solution of (\ref{diracpsi}) and (\ref{diracchi}) is given by
\begin{align}
 u_\psi =& c_\psi \exp{\left(2 \sqrt{|\Delta|}\sqrt{\theta-\theta_+} t \right)} + d_\psi \exp{\left(- 2\sqrt{|\Delta|}\sqrt{\theta-\theta_+} t \right)} \,, \label{spinorsol1}\\
 v_\psi =& \frac{8\ell z}{3\ell^2+6z^2}\left[ (\sqrt{\theta-\theta_0}-\sqrt{\theta-\theta_+})c_\psi \exp{\left( 2\sqrt{|\Delta|}\sqrt{\theta-\theta_+} t \right)} \right. \nonumber\\
 &\left.+ (\sqrt{\theta-\theta_0}+\sqrt{\theta-\theta_+})d_\psi \exp{\left(- 2\sqrt{|\Delta|}\sqrt{\theta-\theta_+} t \right)}\right] \,,\\
 u_\chi =& c_\chi \exp{\left( 2\sqrt{|\Delta|}\sqrt{\theta-\theta_-} t \right)} + d_\chi \exp{\left(- 2\sqrt{|\Delta|}\sqrt{\theta-\theta_-} t \right)} \,,\\
v_\chi =& \frac{8\ell z}{3\ell^2 - 6z^2}\left[ (\sqrt{\theta-\theta_0}-\sqrt{\theta-\theta_-})c_\chi \exp{\left( 2\sqrt{|\Delta|}\sqrt{\theta-\theta_-} t \right)} \right. \nonumber\\
 &\left.+ (\sqrt{\theta-\theta_0}+\sqrt{\theta-\theta_-})d_\chi \exp{\left(- 2\sqrt{|\Delta|}\sqrt{\theta-\theta_-} t \right)}\right] \,,\label{spinorsol4}
\end{align}
where $c_\psi, d_\psi, c_\chi, d_\chi$ are complex integration constants, and
\begin{equation}
 \theta_\pm = \frac{25 \ell^4+100z^4 \pm 36 \ell^2 z^2}{64 \ell^2 z^2}\,.
\end{equation}

\begin{figure}[ht!]
    \centering
        \begin{subfigure}[t]{.5\textwidth}
        \includegraphics[trim={0.05cm 0.7cm 0.05cm 0.5cm},clip,width=1\linewidth]{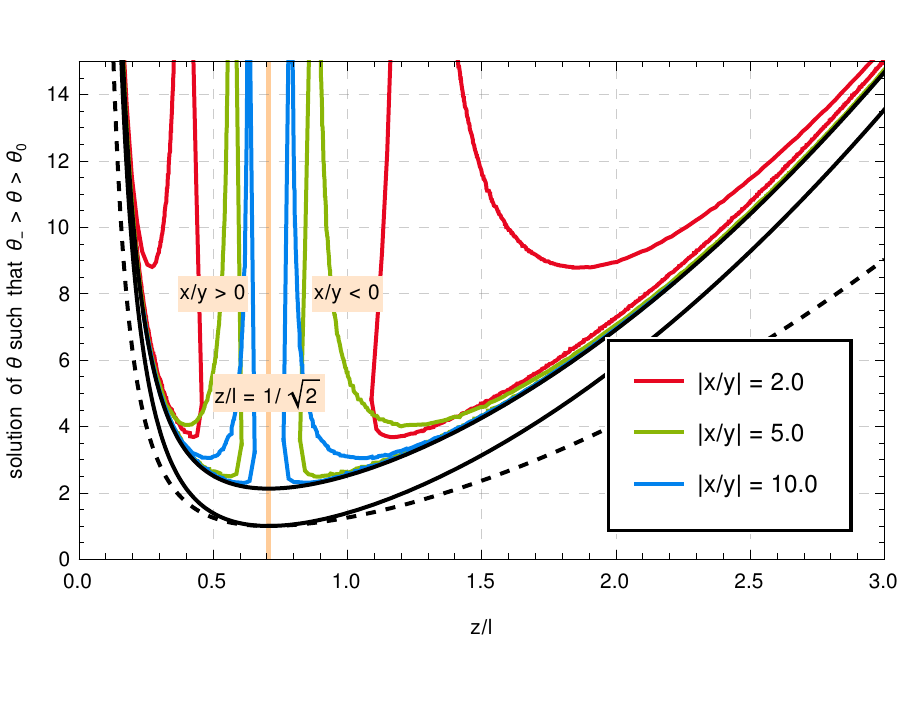}
        \caption{Solutions for $\theta > \theta_+$.}
        \label{fig:region1}
        \end{subfigure}~~\hfill
        \begin{subfigure}[t]{0.5\textwidth}
        \includegraphics[trim={0.05cm 0.7cm 0.05cm 0.5cm},clip,width=1\linewidth]{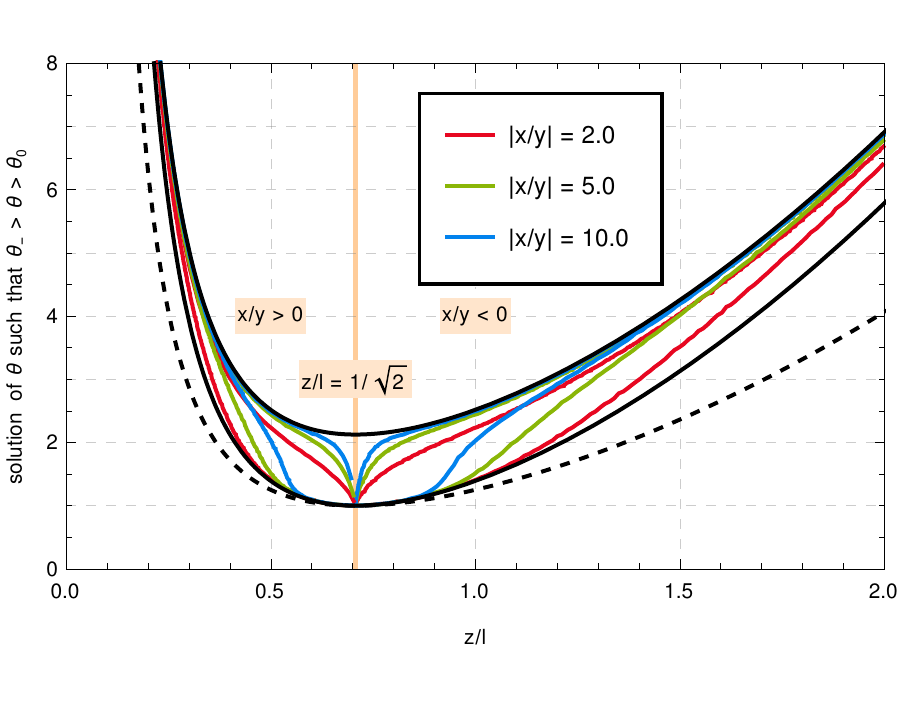}
        \caption{Solutions for $\theta_+ > \theta > \theta_-$.}
        \label{fig:region2}
        \end{subfigure}
        \begin{subfigure}{0.5\textwidth}
        \includegraphics[trim={0.05cm 0.2cm 0.5 0.2cm},clip,width=1\linewidth]{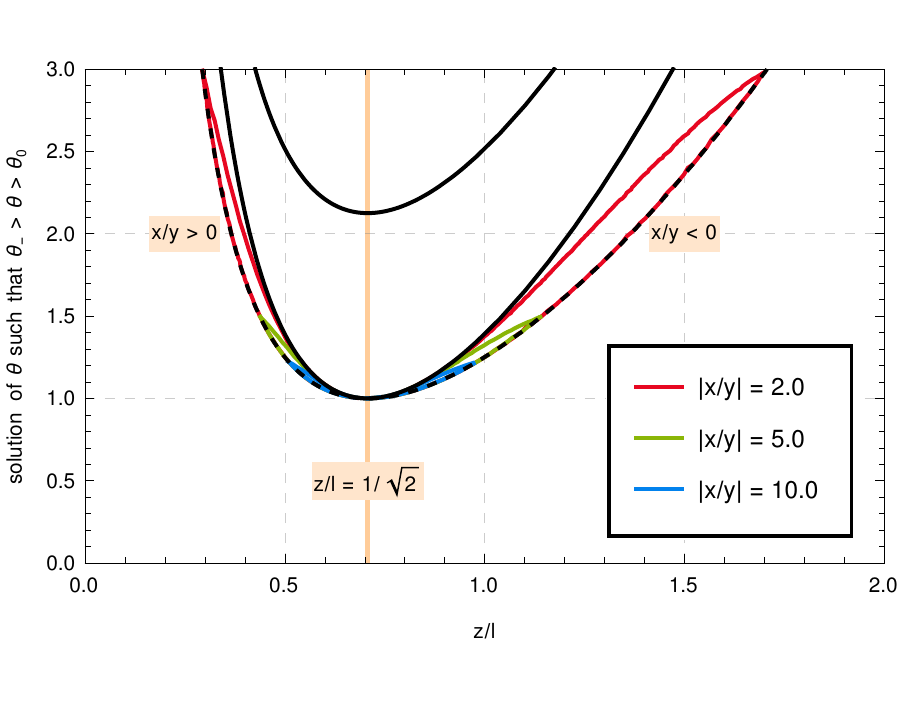}
        \caption{Solutions for $\theta < \theta_-$. The black-dashed line marks the function $\theta_0=\theta_0(z)$, where solutions with $\Delta > 0$ cease to exist.}
        \label{fig:region3}
        \end{subfigure}
        \caption{Black lines correspond to $\theta _+ > \theta \ge \theta_0$, where the equality holds for $z/l=1/\sqrt{2}$.}
\label{figsolutions}
\end{figure}

In order to compute $\Delta$ we need to take into acount the sign of $(\theta-\theta_\pm)$, see fig. \ref{fig:region3}. We will only consider $\theta > \theta_0$, the border case $\theta = \theta_0$ correspond to flat Minkowski. Assuming fixed $\ell$ and $0<z<\infty$ we see that $1 \leq \theta_0  < \infty$, where $\theta_0^\text{(min)}=1$ for $z=1/\sqrt{2}$.  We can also see that $\theta_\pm = (25/16)\theta_0\pm(36/64)$, therefore the following order of parameters is always valid
\begin{equation}
 \theta_+ > \theta_- > \theta_0\,. \label{hierarchy}
\end{equation}

The evaluation of $\Delta$ and $\Phi$ give us,
\begin{align}
 \Delta =& \frac{16\ell z}{3(\ell^2+2z^2)}\left(\sqrt{|\theta - \theta_+|}\ x - \frac{\ell^2+2z^2}{\ell^2-2z^2}\sqrt{|\theta - \theta_-|}\ y\right)\,,\\
 \Phi =& \frac{16\ell z}{3(\ell^2+2z^2)}\left(\sqrt{|\theta - \theta_+|}\ x + \frac{\ell^2+2z^2}{\ell^2-2z^2}\sqrt{|\theta - \theta_-|}\ y\right)\,,
\end{align}
where
\begin{align}
 x =& \begin{cases}
i(\overline{c_\psi}d_\psi - \overline{d_\psi}c_\psi)\,, \quad \text{if } \theta > \theta_+\\
|\overline{c_\psi}|^2 - |\overline{d_\psi}|^2 \,, \quad \text{if } \theta < \theta_+
\end{cases}\,,\\
y =& \begin{cases}
i(\overline{c_\chi}d_\chi - \overline{d_\chi}c_\chi)\,, \quad \text{if } \theta > \theta_-\\
|\overline{c_\chi}|^2 - |\overline{d_\chi}|^2 \,, \quad \text{if } \theta < \theta_-
\end{cases}\,,
\end{align}

The consistency condition can be written as
\begin{equation}
 \theta = \frac{(\ell^2-2z^2)\sqrt{|\theta - \theta_+|}\ x+(\ell^2+2z^2)\sqrt{|\theta - \theta_-|}\ y}{|(\ell^2-2z^2)\sqrt{|\theta - \theta_+|}\ x-(\ell^2+2z^2)\sqrt{|\theta - \theta_-|}\ y|}\,. \label{consistencytheta}
\end{equation}
There are no obstructions to the existence of physical solutions of (\ref{consistencytheta}). For the sake of definiteness let us focus on $\Delta >0$, where solutions can be found for: $x/y>0$ if $z/\ell < 1/\sqrt{2}$, and $x/y<0$ if $z/\ell>1/\sqrt{2}$, see figure \ref{figsolutions}. The case $z/\ell = 1/\sqrt{2}$ is singular and only flat solutions exist.

Numerical solutions to (\ref{consistencytheta}) can be found. At a fixed $z$, and for sufficiently big $|x/y|$, there are two solutions for $\theta > \theta_+$, except at the vertical asymptotes located at
\begin{equation}
 1-\left( \frac{\ell^2+2z^2}{\ell^2-2z^2} \right)^2\left(\frac{y}{x}\right)^2 = 0\,,\label{verticalasymp}
\end{equation}
where only one solution survives, see figure (\ref{fig:region1}). Condition (\ref{verticalasymp}) comes from demanding the vanishing of the coefficient of $\theta$ in the denominator of (\ref{consistencytheta}) after the rationals have been removed. The solutions for $z/\ell >> 1$ have the asymptotic behaviour
\begin{equation}
 \theta = \overline{\theta}_+ + \frac{(9/8) |y|^2}{|x|^2+|y|^2} + O((z/\ell)^{-4})\,, \label{shift}
\end{equation}
where $\overline{\theta}_+ = (100 z^2+ 36)/64$. The asymptotic relation (\ref{shift}) tell us that solutions are located at a constant shift with respect to $\theta_+$ for large $z/\ell$. No solutions are found for $\theta = \theta_+$.

In the region $\theta_- < \theta <\theta_+$, see figure \ref{fig:region2}, there are now two solutions for any value of $z$ such that $z/\ell \ne 1/\sqrt{2}$. One branch in this region is parallel to $\theta_-$ at $z/\ell = 1/\sqrt{2}$ and the other branch is parallel to a vertical line at $z/\ell = 1/\sqrt{2}$ at $z/\ell = 1/\sqrt{2}$. For large $z$ both branches tend to a constant shift with respect to $\theta_+$. No solutions are found at $\theta = \theta_-$, except for $\theta = 1$ but that one is related to a flat solution since it occurs at $z/\ell = 1/\sqrt{2}$ where $\theta_- = \theta_0$. 

In the region $\theta < \theta_-$, see figure \ref{fig:region3}, we found two solutions for sufficiently small $|x/y|$. One solution is perfectly acceptable but the other one lies on top of the $\theta = \theta_0$ line and therefore it corresponds to flat solutions.

Also analytical solutions under appropriate approximations can be found. For illustration let us consider the upper branch in case $\Delta >0$ and $z/\ell \gg 1/\sqrt{2}$, see figure \ref{fig:region1}. To search for solutions in this case we can use the expansion
\begin{equation}
 \theta = \overline{\theta}_+ + \varepsilon + O((z/\ell)^{-4})\,, 
\end{equation}
where $\varepsilon \sim (z/\ell)^0$. The value of the constant $\varepsilon$, was given in (\ref{shift}), can be fixed by demanding the vanishing of the coefficient multiplying the higher order in $z$ of the denominator of (\ref{consistencytheta}). There are similar approximations that can be taken in other limiting cases.

\subsection{Energy condition}\label{energycond}

In three dimensions we define the energy momentum 2-form by
\begin{equation}
 \delta L = \delta e^a \tau_a\,,
\end{equation}
where
\begin{equation}\label{taud}
 \tau_a=\frac{1}{2}\tau_a{}^\mu\epsilon_{\mu bc} e^b e^c\,.
\end{equation}

The variation of (\ref{spinorlag}) with respect to $e^a$ give us
\begin{align}
 \delta_e L=& \delta e^a \kappa \left[
 \epsilon_{abc}\psibar(e^b\gamma^c D+\overleftarrow{D}e^b\gamma^c)\psi
 -\epsilon_{abc}\chibar(e^b\gamma^c D+\overleftarrow{D}e^b\gamma^c)\chi \nonumber \right.\\
&\left.-2T_a\Delta-\frac{1}{\ell}\epsilon_{abc}e^bf^c \Phi-i\epsilon_{abc}e^b B_I(\psibar\gamma^c\sigma^I\psi + \chibar\gamma^c\sigma^I\chi)+e_a d\Delta\right]\,,
\end{align}
plus a boundary term (given by $-\kappa d(\delta e^a e_a \Delta)$). Therefore

\begin{align}
 \tau_{ab} =& \kappa \left[ \eta_{ab} \psibar (\gamma^c D_c -\lD_c\gamma^c ) \psi - \half \psibar\left( \gamma_a D_b +\gamma_b D_a \right)\psi + \half \psibar\left( \lD_b\gamma_a  +\lD_a\gamma_b \right)\psi\right.\nonumber\\
 &-\eta_{ab} \chibar (\gamma^c D_c -\lD_c\gamma^c ) \chi + \half \chibar\left( \gamma_a D_b +\gamma_b D_a \right)\chi - \half \chibar\left( \lD_b\gamma_a  +\lD_a\gamma_b \right)\chi \nonumber\\
 &\left.+\frac{2hl}{z^2}\left(\Delta + \frac{z^2}{l^2}\Phi\right)\eta_{ab}\right]\,.
\end{align}

The evaluation for the solution (\ref{spinorsol1})-(\ref{spinorsol4}) give us
\begin{align}
 \tau_{00} \equiv \rho =& -(\rho_\Delta+2\rho_\Phi) \,,\label{tau00}\\
 \tau_{02} \equiv P_2 =& -2(z/\ell)\sqrt{\theta-\theta_0} \ \rho_\Delta\,,\\
  \tau_{11} =\tau_{22} \equiv p =& \frac{5}{2} ( \rho_\Delta + 2\rho_\Phi )\,,\label{tau11}
\end{align}
where
\begin{equation}
 \rho_\Delta=\kappa {(z/\ell)}^{-1}\sqrt{|\Delta|}\Delta \,, \quad \rho_\Phi = \kappa (z/\ell)\sqrt{|\Delta|} \Phi \,.
\end{equation}

\begin{figure}[ht!]
    \centering
        \begin{subfigure}[t]{1.0\textwidth}
        \includegraphics[trim={0.05cm 0.2cm 0cm 0.05cm},clip,width=.5\linewidth]{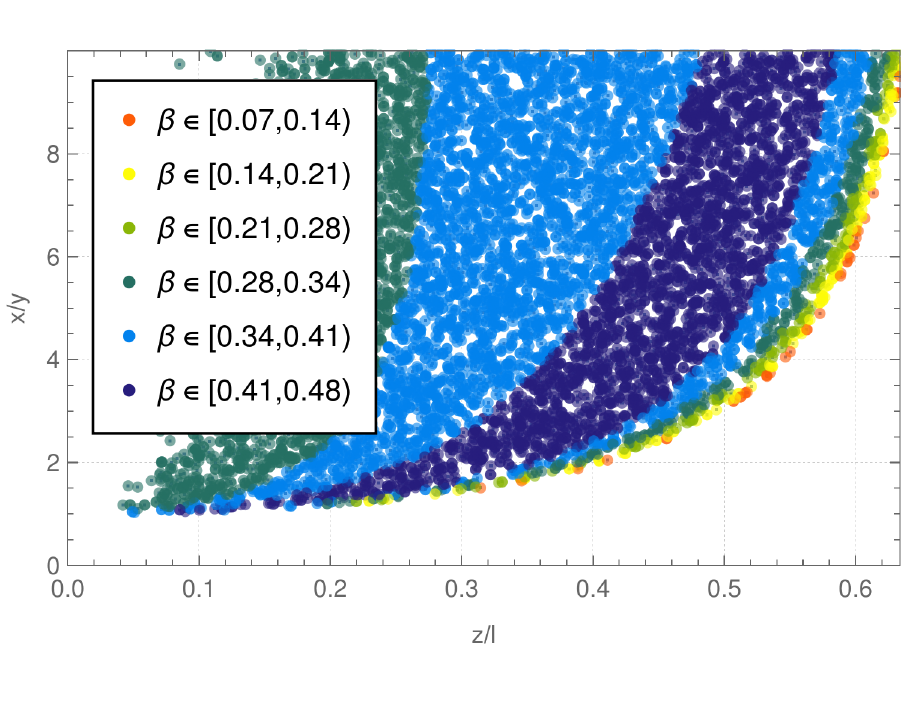}
        \includegraphics[trim={0.05cm 0.2cm 0cm 0.05cm},clip,width=.5\linewidth]{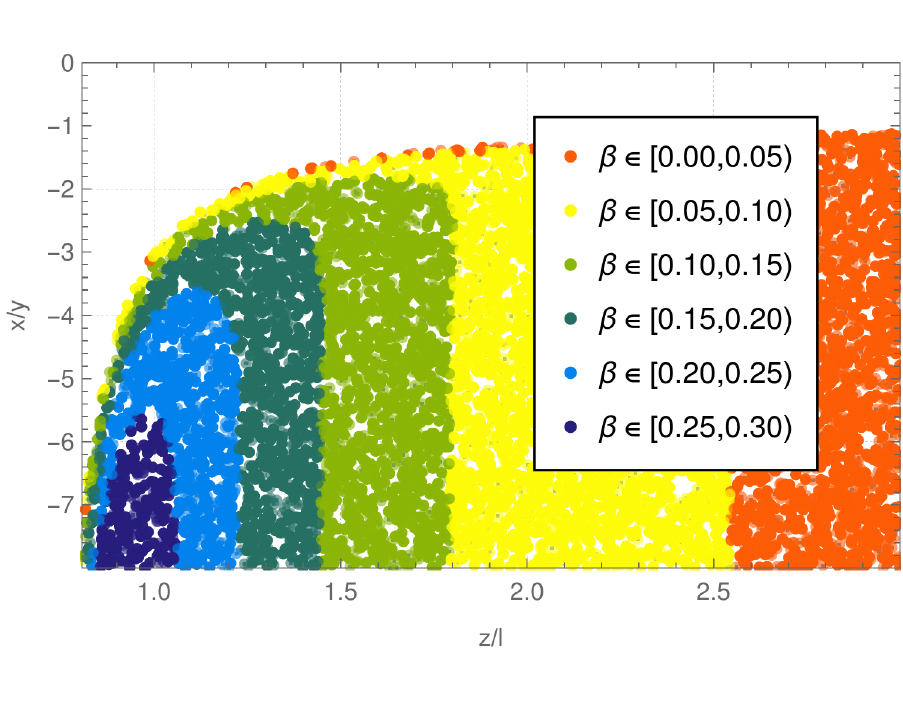}
        \caption{Solutions for $\theta > \theta_+$. The higher rapidities are obtained for this class of solutions, specially when $z/\ell < 1/\sqrt{2}$. This case is not smoothly connected to the vacuum $x=0$ and $y=0$ case.}
        \label{fig:region1beta}
        \end{subfigure}
        \begin{subfigure}[t]{1.0\textwidth}
        \includegraphics[trim={0.05cm 0.2cm 0cm 0.05cm},clip,width=.5\linewidth]{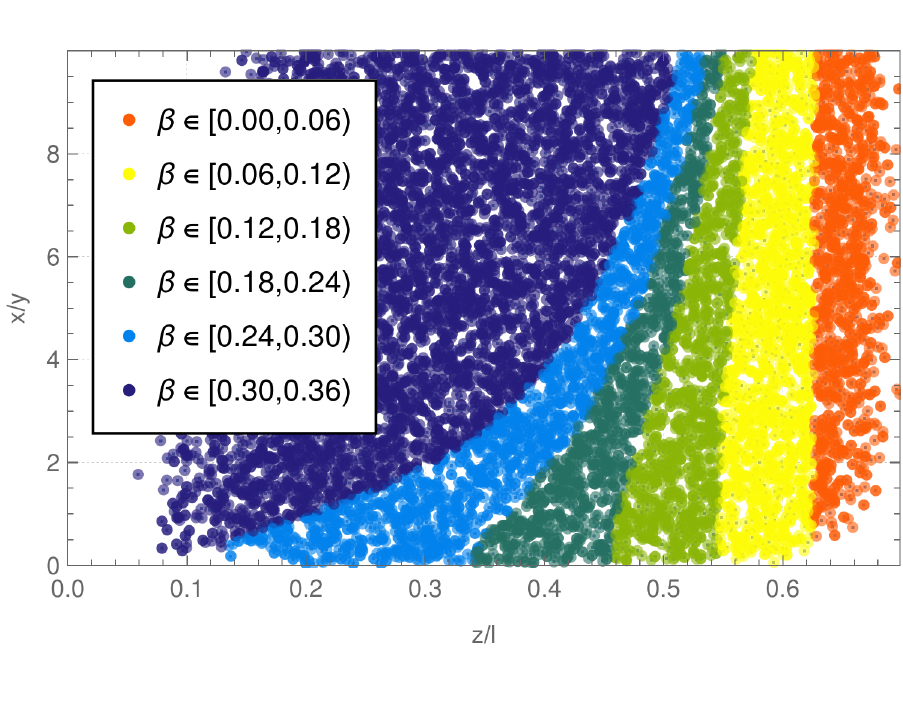}
        \includegraphics[trim={0.05cm 0.2cm 0cm 0.05cm},clip,width=.5\linewidth]{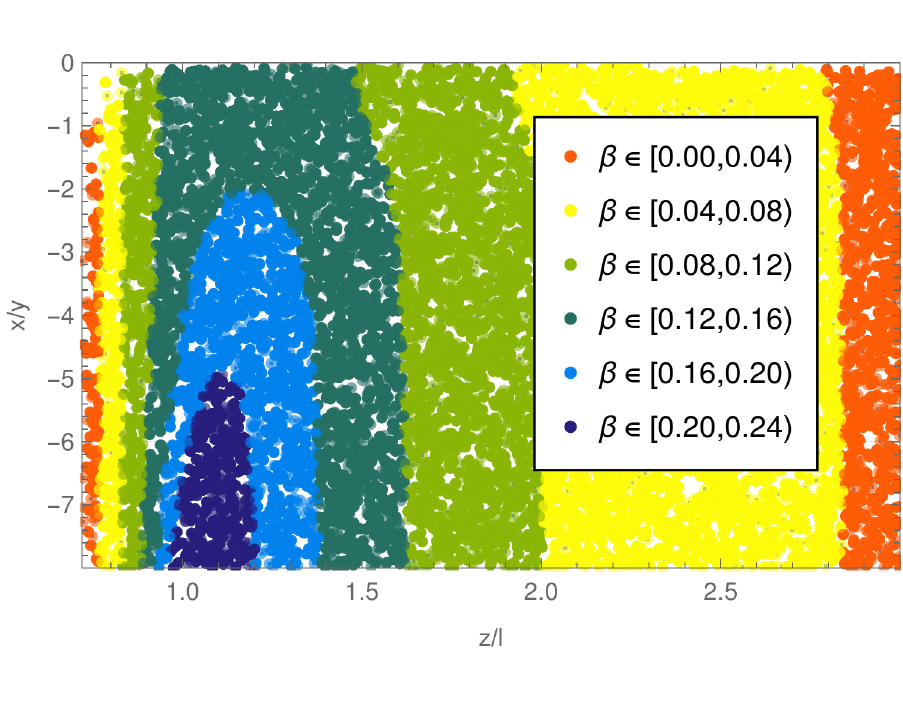}
        \caption{Solutions for $\theta_+ > \theta > \theta_-$. This case can be smoothly deformed to the vacuum $x=0$ and $y=0$ case and the torsionless case $z \rightarrow \infty$.}
        \label{fig:region2beta}
        \end{subfigure}
        \begin{subfigure}[t]{1\textwidth}
        \includegraphics[trim={0.05cm 0.2cm 0cm 0.05cm},clip,width=.5\linewidth]{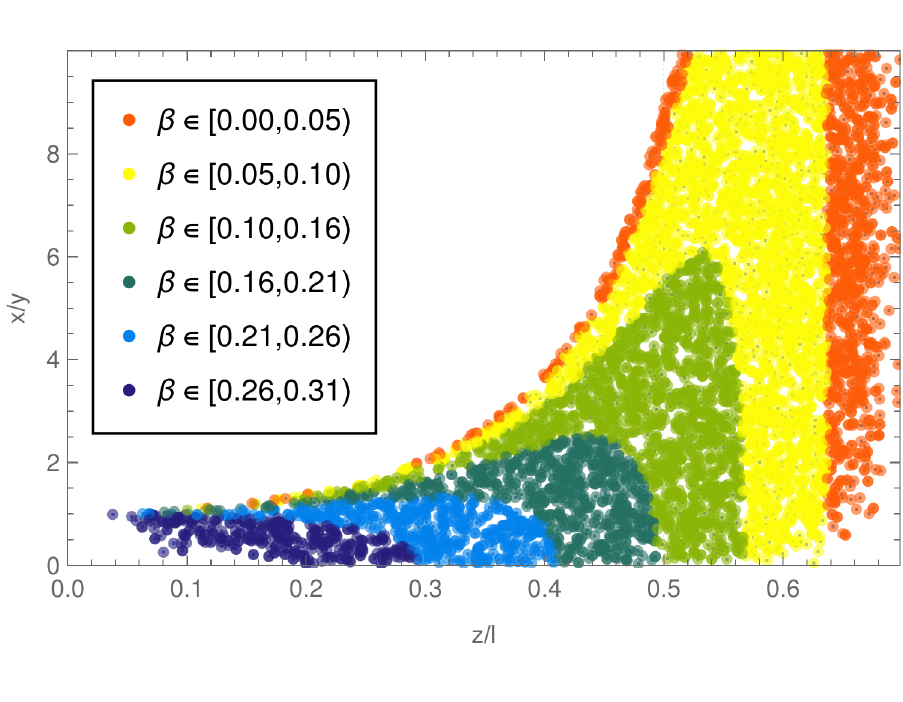}
        \includegraphics[trim={0.05cm 0.2cm 0cm 0.05cm},clip,width=.5\linewidth]{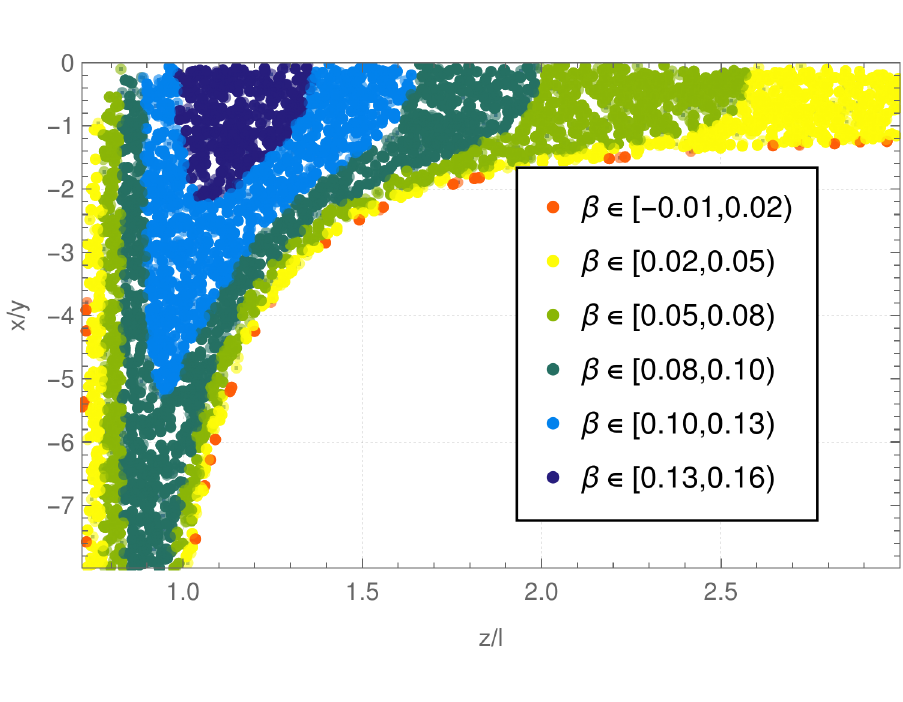}
        \caption{Solutions for $\theta < \theta_-$. This case can also be smoothly deformed to the vacuum $x=0$ and $y=0$ case and the torsionless case $z \rightarrow \infty$.}
        \label{fig:region3beta}
        \end{subfigure}
        \caption{Rapidity parameter of the fluid.}
\label{betafig}
\end{figure}

Relations (\ref{tau00}) - (\ref{tau11}) show that neither the weak or the strong energy conditions are satisfied, however such patologies are traced back to the fact that the backrogund geometry is adS. Leaving aside the issues with the adS background we can see a striking simplicity: the fluid corresponds to the energy momentum tensor of a fluid with non-isotropic stress where the fluid is in a frame of constant speed in the 2-direction such that the anisotropic pressures are equated (see ref. \cite{Krisch:2001ay}). The tensor can be written as a mixture of matter and vacuum energy
\begin{equation}
 \tau_{ab} = \begin{pmatrix}
\rho_\text{fluid} & 0 & P_2\\
0 & 0 & 0\\
P_2 & 0 & 0
\end{pmatrix}
-\rho_\text{vac} \eta_{ab}\,,
\end{equation}
where $\rho_\text{vac}=-\tfrac{5}{2}(\rho_\Delta+2\rho_\Phi)$, see reference for a general study of a two component fluid \cite{Letelier:1980mxb}. The geometric origin of the solution of the present paper means that $\rho_\text{fluid}$ is, however, not independent from $\rho_\text{vac}$: $\rho_\text{fluid}=-\tfrac{3}{5}\rho_\text{vac}$. Also note in passing that $\rho_\text{fluid} > 0$. The condition for non-superluminality is given by $2|P_2| < |\rho+p|$, where for the present solution it can be written as
\begin{equation}
 1 < \frac{3|1+2(z/\ell)^2\theta|}{8(z/\ell)|\sqrt{\theta-\theta_0}|}\,. \label{notsulcond}
\end{equation}
Solutions are such that for $(z/\ell) \gg 1$ we have $\theta \sim \theta_+ \sim (5z/(4\ell))^2$ and $\theta-\theta_0 \sim ((5/4)^2-1)(z/\ell)^2$, and therefore condition (\ref{notsulcond}) is satisfied. Around $(z/\ell) \sim 1/\sqrt{2}$ we can use $\theta \sim \theta_0 + \varepsilon$ for a small parameter $\varepsilon$, and then it is easy to see that (\ref{notsulcond}) is also satisfied. Numerical analysis also confirms no violations of (\ref{notsulcond}) for physical solutions of $\theta$ ($\Delta > 0$). The speed of the fluid is characterized by the rapidity parameter $\beta = \tanh \lambda$, where $\lambda = \frac{1}{4}\ln \frac{\rho+p-2P}{\rho+p+2P}$. Such transformation corresponds to a Lorentz boost in the 2-direction that acts actively on the energy momentum $\tau_{ab}$. We evaluated numerically those values, and the results show that the rapidity varies from zero to a fraction of the speed of light, see figure \ref{betafig}.

\begin{figure}[ht!]
    \centering
    \pgfmathsetlength{\imagewidth}{\linewidth}%
    \pgfmathsetlength{\imagescale}{\imagewidth/524}%
    \begin{tikzpicture}[x=\imagescale,y=-\imagescale]
        \node[anchor=north west] at (0,0) {\includegraphics[width=0.7\imagewidth]{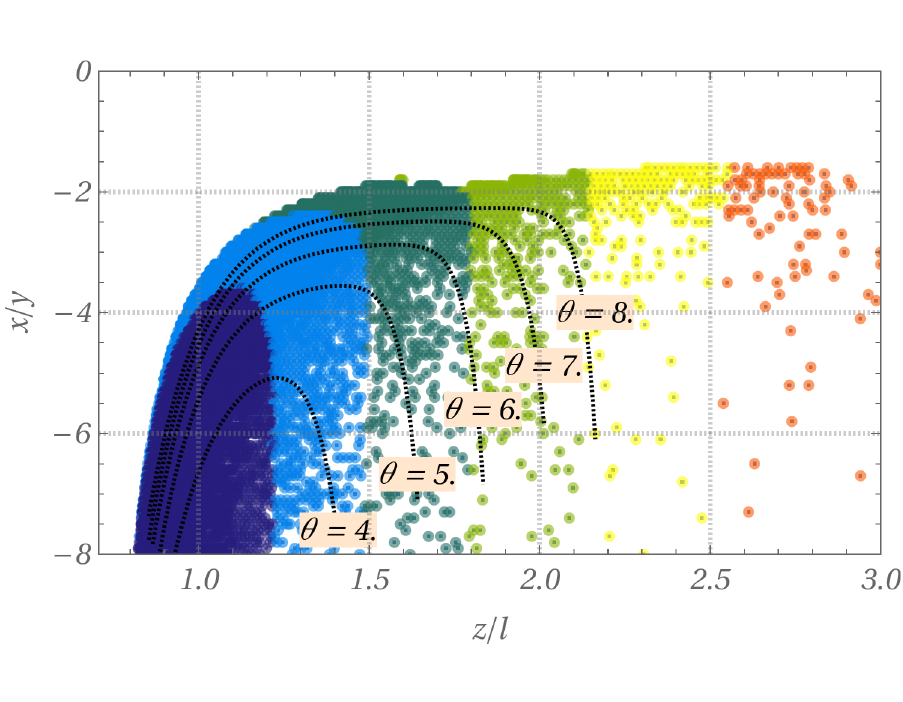}};
        \node[anchor=north west] at (270,90) {\includegraphics[width=0.15\imagewidth]{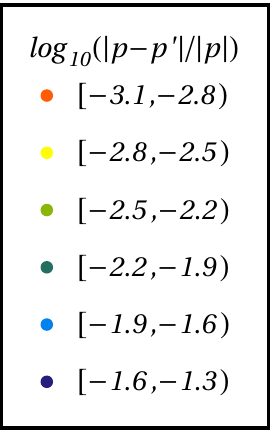}};
    \end{tikzpicture}
    \caption{Pressure anisotropy for solutions such that $\theta > \theta_+$. Here we focus on the upper branch of figure \ref{fig:region1}. Dashed lines serve as reference to visualize the relation between pressure anisotropy and $\theta$. The diminishing density of points towards high values of $z$ is an artifact of the numerical method to obtain the zeros of the multiple branches of eq. (\ref{consistencytheta}).}
\label{figpressure}
\end{figure}

We also computed the pressure anisotropy by going to the rest frame where $P'=0$, $p_1 = p$ and $p_2 = p'$, where
\begin{equation}
 p' = (\rho + p) \cosh (\lambda)^2 + 2 P \cosh(\lambda)\sinh(\lambda) - \rho\,.
\end{equation}
The results can be summarized by defining the anisotropy coefficient $r_p\equiv|p-p'|/|p|$. For values of the spinor parameter $x/y\sim (-2,-8)$ and $z/\ell\sim (1/\sqrt{2},3)$ we obtain that the values of $r_p$ are contained within the range $(10^{-1},10^{-3})$, see figure \ref{figpressure}.

\section{Conclusions}\label{sec:conclu}\label{sec:conclu}

In this paper we studied analytical solutions for a FRW expanding universe with a gravitating spinor for a three dimensional gravity model with matter (spin 1/2) in the adjoint representation of the superalgebra. We found a fluid, that in the rest frame, has anisotropic pressure.

As a future work it would be interesting to explore the existence of similar self-gravitating spinor configurations in Dirac materials \cite{Acquaviva:2022yiq} or planar QED \cite{Dudal:2018mms}.

The spinor solution is not supersymmetric nor can be deformed to the vacuum case $\theta \rightarrow 0$ without taking an improper limit of the other parameters, $z\rightarrow \infty$, and therefore we conclude that generically it cannot be obtained by the perturbative process of ``gauge completion'' defined by Gentile et al \cite{Gentile:2012tu}. It may occurs that for the special cases $\theta = \theta_0$ that the spinor exact solution could be obtained by a perturbative series, since for such a case the associated geometry is flat and the spinor loses half of the degrees of freedom.

The lift of the 2+1 BTZ solution to 3+1 black string was done by Lemos and Zanchin \cite{Lemos:1995cp}. The negative density in this case can be physically motivated from the cosmological constant. It will be of interest to study the possibility of lifting the solutions found in this paper to apply the Geroch transform and then project back down to 2+1 to examine the effects on the cosmological fluid.

It is also interesting to look for more complex configurations where more of the fermionic bilinears are nontrivial.

\section{Acknowledgements}

P. A. acknowledges MINEDUC-UA project ANT 1755 and Semillero de Investigaci\'on project SEM18-02 from Universidad de Antofagasta, Chile. P. A. also acknowledge the hospitality at the Politecnico de Torino, where part of this research was carried out, and the illuminating discussions with Jorge Zanelli, Cristobal Corral, Laura Andrianopoli, Bianca Cerchiai, Lucrezia Ravera and Marco Astorino. J. O. acknowledges the support of the ANID doctoral fellowship 21220369 and the the PhD program Doctorado en F\'isica menci\'on en F\'isica Matem\'atica de la Universidad de Antofagasta for continuous support and encouragement.

\begin{appendices}

\section{$su(2|2)$ representation}\label{apprep}

Let us consider the following representation of $su(2|2)$
\begin{eqnarray}
&\mathbb{J}_a =\left[\begin{array}{c|c}
\frac{1}{2}\gamma_a &  0_{2\times2}\\[0.5em] \hline
0_{2\times2} & 0_{2\times2} \\
\end{array}\right]\,, \quad \text{or} \quad (\mathbb{J}_a)^A_{\ B}=\frac{1}{2}(\gamma_a)^A_{\ B}\,,&\\
&\mathbb{T}_{I} =\left[\begin{array}{c|c}
0_{2\times2} &  0_{2\times2}\\[0.5em] \hline
0_{2\times2} & -\frac{i}{2} (\sigma_I)^t \\
\end{array}\right]\,, \quad \text{or} \quad (\mathbb{T}_I)^A_{\ B}=-\frac{i}{2}(\sigma_I^t)^A{}_B\,,&\\
&(\mathbb{Q}^\alpha_i)^A_{\ B}=\left[\begin{array}{c|c}
0_{2\times2} & 0_{2\times 2}\\ [0.5em] \hline
\delta^A_i \delta^\alpha_B & 0_{2\times2}
\end{array}\right]\,,&\\
&(\overline{\mathbb{Q}}_\alpha^i)^A_{\ B}=\left[\begin{array}{c|c}
0_{2\times2} & \delta^A_\alpha \delta^i_B\\ [0.5em] \hline
0_{2\times 2} & 0_{2\times2}
\end{array}\right]\,,&\\
&\mathbb{Z}^A_{\ B}=\left[\begin{array}{c|c}
\frac{i}{2}\delta^\alpha_\beta &0_{2\times2}\\ [0.5em] \hline
0_{2\times2} & \frac{i}{2}\delta^i_j\end{array}\right]=\frac{i}{2}(\delta^A_\alpha \delta^\alpha_B+\delta^A_i \delta^i_B)\,,&
\end{eqnarray}
The $\gamma$-matrices are in a $2\times 2$ spinorial-representation with indices $\alpha, \beta=1,2$. The indices of the tangent space $a,b=0,1,2$. Indices in the adjoint representation of $SU(2)$ take values $I,J=1,2,3$ and indices in the fundamental take values $i,j=1,2$. We chose the upper-left block of the representation for spinor indexes and lower-right block for the indexes in the fundamental representation of $SU(2)$, so the split $A=(\alpha,i)$.

The $\gamma$-matrices satisfy $\{\gamma^a,\gamma^b\}=2 \eta^{ab}$, where the metric $\eta$ is given by $\eta=\mathrm{diag}(-,+,+)$, and the spinorial indexes are often omitted. We have two complex spinors $\psi^\alpha_i$. The Pauli matrices satisfy $[\sigma_I,\sigma_J]=2i\epsilon_{IJ}^{\ \ \ K} \sigma_K$.

Using the fact that in three dimensions the $\gamma$-matrices satisfy $\gamma_a \gamma_b \gamma_c=\epsilon_{abc}$, where $\epsilon_{012}=1=-\epsilon^{abc}$, it is checked that $[\gamma_a,\gamma_b]=2 \epsilon_{ab}^{\ \ \ c}\gamma_c$ so generators $\mathbb{J}_a$ form a $so(2,1) \cong su(2)$ algebra,
\begin{equation}
[\mathbb{J}_a,\mathbb{J}_b]=\epsilon_{ab}^{\ \ \ c}\mathbb{J}_c\,.
\end{equation}
The relation to the conventional double index Lorentz generators is given by
\begin{equation}
 J^a = \half \epsilon^a{}_{bc} J^{bc} \,, \quad J_{ab}=-\epsilon_{abc} J^c\,,
\end{equation}
and $\Sigma_{ab}=(1/2)[\gamma_a,\gamma_b]$.

For the internal generators we have the $su(2)$ algebra
\begin{equation}
 [\mathbb{T}_I,\mathbb{T}_J]=\epsilon_{IJ}^{\ \ \ K}\mathbb{T}_K\,,
\end{equation}
and they are anti-Hermitian $\mathbb{T}_I^\dag=-\mathbb{T}_I$.

Inlcuding the supercharges all the (anti-)commutators close in a $su(2|2)$ superalgebra
\begin{eqnarray}
&[\mathbb{J}_a,\overline{\mathbb{Q}}_\alpha^i]=\frac{1}{2}\overline{\mathbb{Q}}_\beta^i(\gamma_a)^\beta_{\ \alpha}\,, \quad [\mathbb{J}_a,\mathbb{Q}^\alpha_i]=-\frac{1}{2}(\gamma_a)^\alpha_{\ \beta}\mathbb{Q}^\beta_i\,,&\\
&[\mathbb{T}_I,\overline{\mathbb{Q}}_\alpha^i]=-\frac{i}{2}\overline{\mathbb{Q}}_\alpha^j (\sigma_I)_j^{\ i}\,, \quad [\mathbb{T}_I,\mathbb{Q}^\alpha_i]=\frac{i}{2}(\sigma_I)_i^{\ j}\mathbb{Q}^\alpha_j\,,&\\
&\{\mathbb{Q}^\alpha_i,\overline{\mathbb{Q}}_\beta^j\}=\delta^j_i(\gamma^a)^\alpha_{\ \beta} \mathbb{J}_a-i\delta^\alpha_\beta(\sigma^I)_i^{\ j}\mathbb{T}_I-i\delta^j_i\delta^\alpha_\beta \mathbb{Z}\,.&
\end{eqnarray}
\subsection*{Super-traces:}
The graduation operator is given by
\begin{equation}
\mathcal{G}^A_{\ B} = \delta^A_{\alpha}\delta^\alpha_{\ B}-\delta^{A}_{i}\delta^i_{\ B}\,,
\end{equation}
it classifies generators in bosonic $B=\{\mathbb{J}_a,\mathbb{T}_I,\mathbb{Z}\}$ or fermionic $F=\{\mathbb{Q}^\alpha_i,\overline{\mathbb{Q}}^i_\alpha\}$, by $[B,\mathcal{G}]=0=\{F,\mathcal{G}\}$, and it squares to one $\mathcal{G}^2=1$. With the graduation operator we can define an invariant supertrace
\begin{equation}
\langle G\rangle  \equiv \text{tr}(\mathcal{G} G)=0\,.
\end{equation}
This grants that
\begin{equation}
 \langle B_1 B_2\rangle =\langle B_2 B_1\rangle \,, \quad \langle B F\rangle =\langle F B \rangle \,, \quad \langle F_1 F_2\rangle =-\langle F_2 F_1\rangle \,.
\end{equation}

All the generators $G$ in the representation are supertraceless
\begin{equation}
\langle G\rangle=0\,, \quad G=\{\mathbb{J}_a,\mathbb{T}_I,\mathbb{Z},\mathbb{Q}^\alpha_i,\overline{\mathbb{Q}}^i_\alpha\}\,,
\end{equation}
but some quadratic combinations give nontrivial traces
\begin{eqnarray}
&\langle \mathbb{J}_a \mathbb{J}_b \rangle=\frac{1}{2}\eta_{ab}\,, \qquad \langle \mathbb{T}_I \mathbb{T}_J \rangle=\frac{1}{2}\delta_{IJ}\,,&\\
&\langle \mathbb{Q}^\alpha_i \overline{\mathbb{Q}}^j_\beta\rangle=-\delta^\alpha_\beta \delta^j_i=-\langle \overline{\mathbb{Q}}^j_\beta \mathbb{Q}^\alpha_i\rangle\,.&
\end{eqnarray}

\end{appendices}

\bibliographystyle{ieeetr}
\bibliography{paper.bib}

\end{document}